# Effect of Magnetisation of Bulk Superconductors on Permanent Magnets for Reversible Braking Applications

Nithin Goona[1]*, P. S. Reddy[1], S. Sashidhar[2]
[1]National Institute of Technology Goa, Farmagudi, Ponda, Goa, India
[2]Indian Institute of Technology Goa, Farmagudi, Ponda, Goa, India
*email: nithingoona@nitgoa.ac.in

**Abstract:** Magnetisation of superconductors are often measured under isothermal and uniform magnetic fields. Magnetisation measurements in an unrestricted thermodynamic state and in non uniform magnetic fields naturally emerge from permanent magnets in comparison with size of superconducting bulk are needed for practical applications. A stationary Type II superconductor bulk sample is magnetised (FC and ZFC) at liquid nitrogen temperatures in the presence of a stationary permanent magnet. Magnetic field produced by magnetised superconductor and permanent magnet as a function of distance are measured along a straight line, a supposed path of another moving permanent magnet. A moving magnet is introduced and moved along the same straight line. Force acting on moving magnet as a function of distance are measured due to magnetisation in superconductor and stationary magnet. Both field and force measurements are quasi static. These experimental results combined with simulation results show insight in to the transfer of energy stored in magnetisation to thermal energy and the effect of rate of this energy transfer in AC response of superconductors which are otherwise not possible with isothermal studies. A possible application of reversible braking with bulk superconductors for motors and generators is explored. Future similar measurements with Type I superconductor would provide more insights in to this study.

**1. Introduction:** Meissner effect is the characteristic phenomenon of superconductors independent of being perfect conductors. Fig. 1 [1] shows a superconductor cylinder surrounded by tiny compass needles placed in uniform magnetic field. When the cylinder is cooled below the critical temperature by placing in liquid Helium, magnetic field is expelled as shown in Fig. 2 [1].

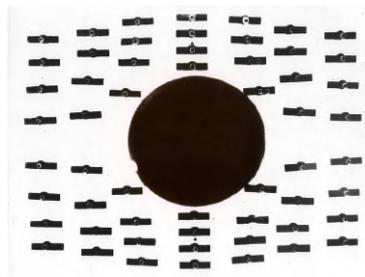 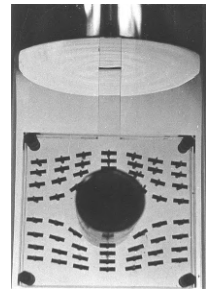

Fig. 1  Magnetic field applied on superconductor      Fig. 2  Meissner effect

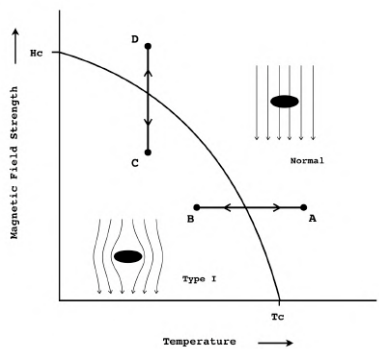 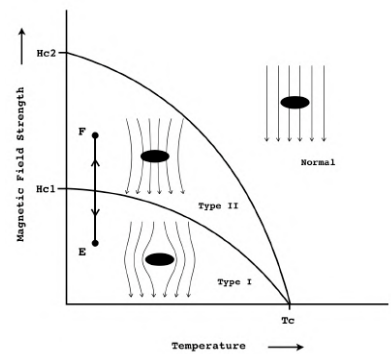

Fig. 3  Phase diagram of Type I superconductor      Fig. 4  Phase diagram of Type II superconductor

The magnetic phase diagrams [2] of Type I and Type II superconductors are shown in Fig. 3 and Fig. 4 respectively showing Meissner and pinning effects. The phase transitions A→B, C→D in Fig. 3, and E→F in Fig. 4 are known to be reversible [3,4,5,6]. The force on permanent magnets due to magnetisation of Type II superconductors [7] is known. The current study focuses on the force on permanent magnet due to magnetisation of Type II superconductor in the presence of another stationary permanent magnet. Magnetic fields around magnet and Type II superconductor bulk are measured. Force on a permanent magnet due to magnetisation of a stationary magnet and Type II superconductor bulk. The method for measuring force is also validated. All the measurements are quasi-static. Experimental results are extended to simulations with Type I and Type II models to realise reversible braking.

**2. Magnetic Field Experimental Measurements:** Fig. 5 shows schematic for field measurements where magnet is placed next to Type II superconductor BSCCO bulk, fields are measured at discrete positions on Line L separated by distance $a$ from the magnet. The extension length of superconductor from magnet $b$ is set constant to 3 mm through out all the experiments. Experiment setup for field measurement is shown in Fig. 6. Lakeshore F71 Multi-axis teslameter with 3 axis probe is used for field measurements. Fig. 7 shows magnet adjacent to superconductor in a container and 3-axis Hall probe.

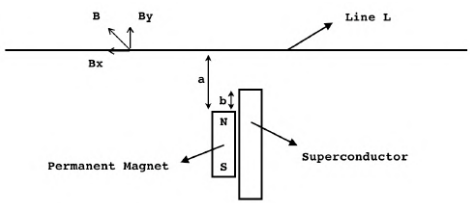
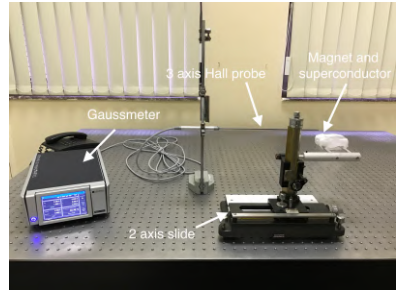
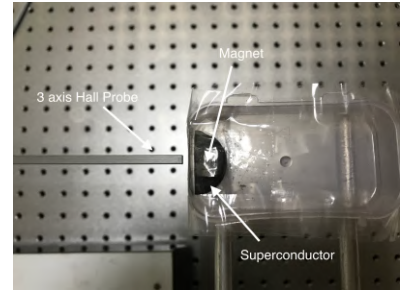

Fig. 5  Schematic for field measurements     Fig. 6  Experimental setup for field     Fig. 7  Positioning of Hall sensor

Fig. 8 and Fig. 9 show x and y component of magnetic fields vs distance respectively with $a = 9$ mm, red and blue curves indicate measurements when superconductor is hot (room temperature) and cold (liquid Nitrogen temperature) respectively. Fig. 10 and Fig. 11 are similar curves to Fig. 8 and Fig. 9 respectively with $a = 6$ mm.

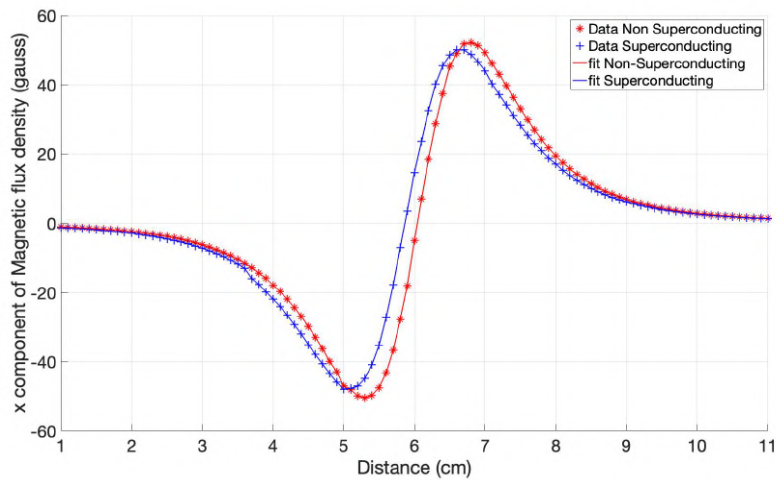

Fig. 8  x component of magnetic field vs distance with $a = 9$ mm

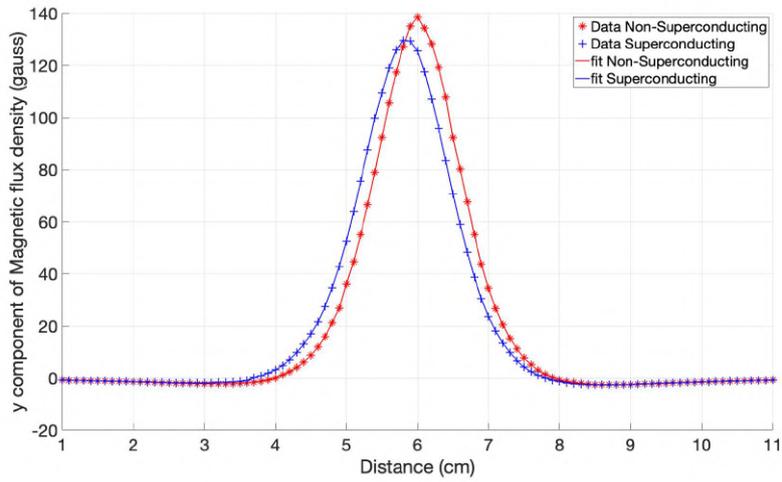

Fig. 9  y component of magnetic field vs distance with *a* = 9 mm

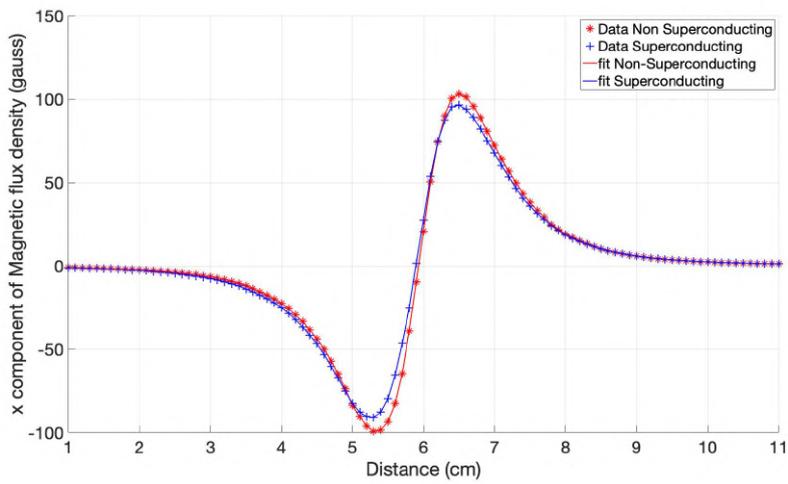

Fig. 10  x component of magnetic field vs distance with *a* = 6 mm

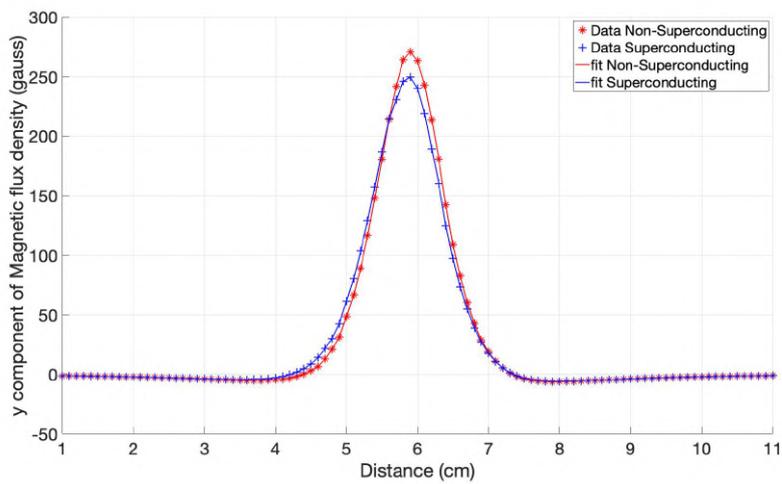

Fig. 11  y component of magnetic field vs distance with *a* = 6 mm

From Fig. 8 and Fig. 10; and from Fig. 9 and Fig. 11, it can be observed that there is a shift of peaks along x direction.

**3. Validation of Force Measurements:** When a magnet is introduced and forces on the magnet are to be measured, a force measurement setup which measures force accurately in one direction independent of forces acting in other directions. Shimadzu UW6200H weighing scale is used and is validated for its accuracy in compensation. Fig. 12 shows schematic for force measurement validation. A circular disc of soft ferromagnetic material is subjected to force due to a permanent magnet which can be moved along a circular arc with magnetisation always in radial direction, that magnitude of force is constant when *d* is maintained constant. Fig. 13 shows experimental setup for force validation. Fig. 14 shows a magnet inside aluminium casing where the position of magnet can be adjusted by tightening the screw. Fig. 15 and Fig. 16 shows vertical component of force vs angle i.e orientation of magnet and deviation curves respectively for different values of *d* = 10 and 6 mm, blue and red respectively.

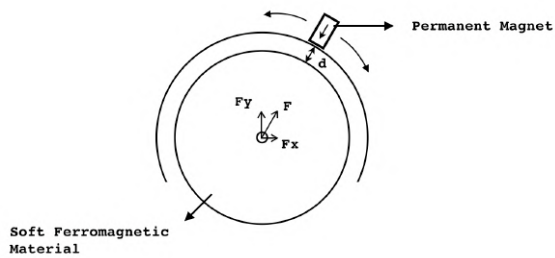 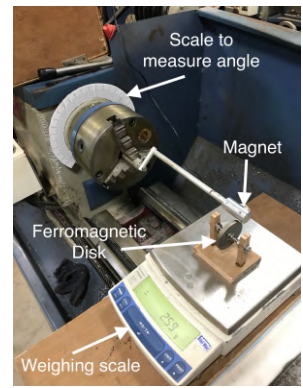

Fig. 12  Schematic for validation of force measurements    Fig. 13  Experimental setup

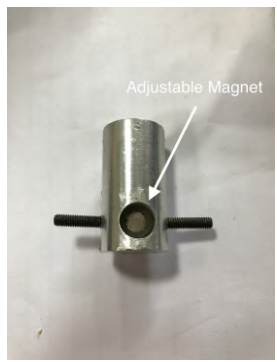

Fig. 14 Magnet-Aluminium enclosure

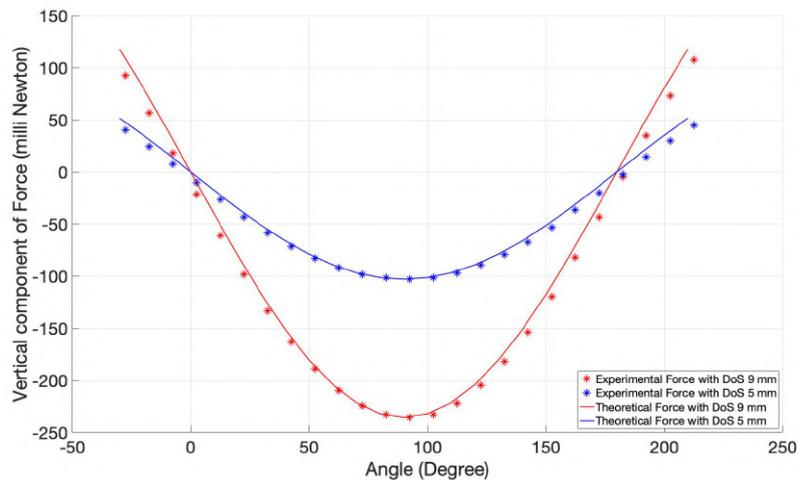

Fig. 15  Vertical Force vs angle from horizontal curves

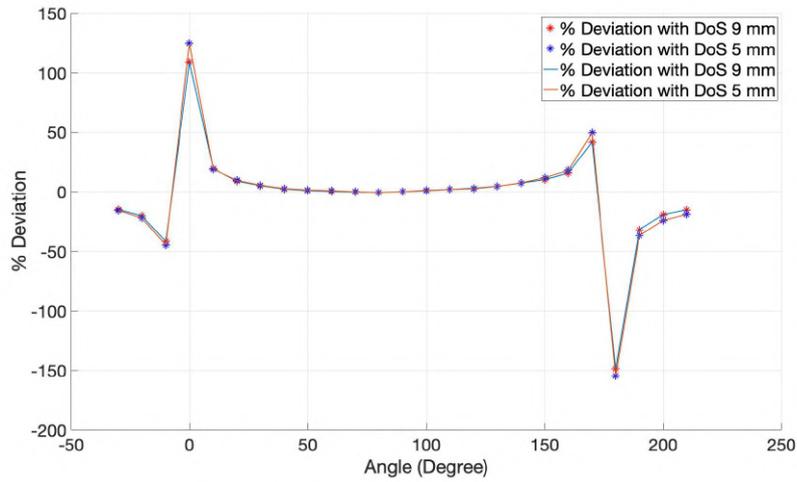

Fig. 16  Deviation vs angle

**4. Force on Magnet Experimental Measurements:** Schematic for Force measurement is shown in Fig. 17 where a moving magnet is moved along the Line L in the presence of a stationary permanent magnet and superconductor. Experimental setup for force measurement is shown in Fig. 18 where moving magnet is placed on weighing scale.

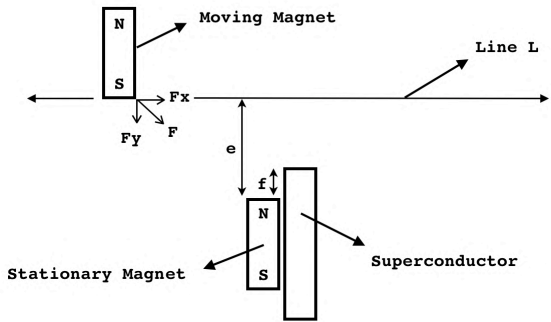
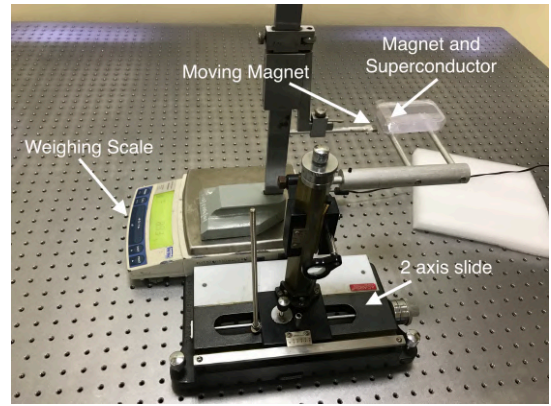

Fig. 17  Schematic for force measurements     Fig. 18  Experimental setup for field

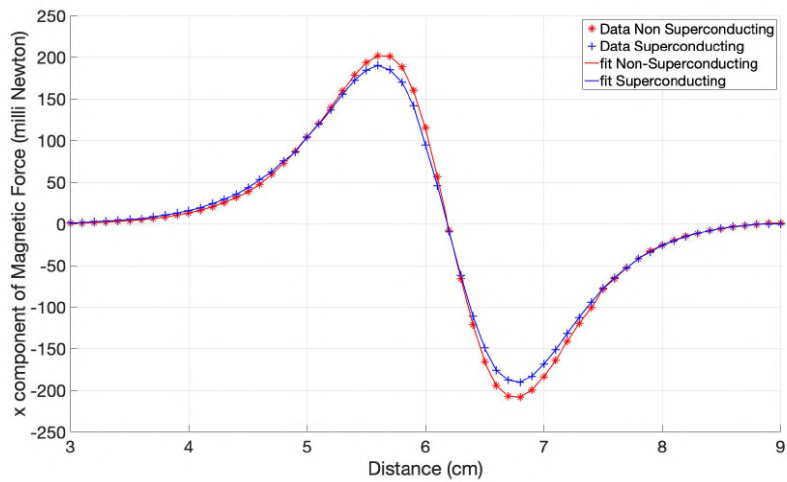

Fig. 19  x component of force vs distance with e = 9 mm

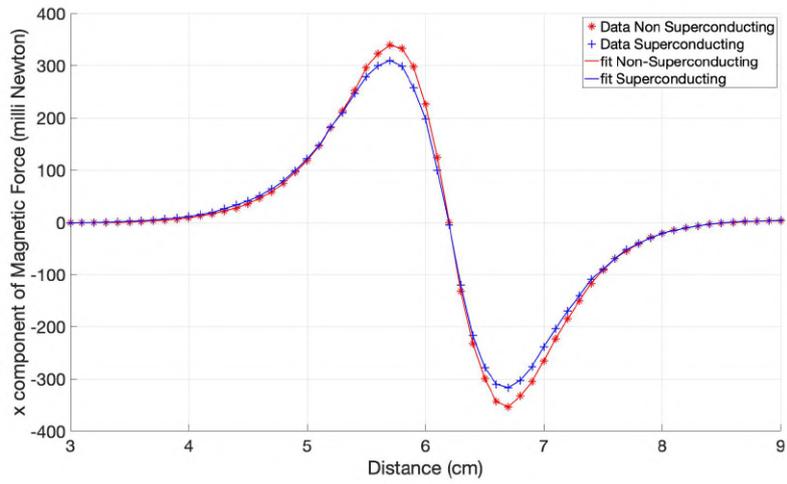

Fig. 20  x component of force vs distance with e = 6 mm

Fig. 19 and Fig. 20 show x component of magnetic fields vs distance with *e* = 9 mm and 6 mm respectively.

**5. Simulation Results:** Fig. 21 and Fig. 22 show simulation results of x and y components of magnetic field  vs distance curves respectively using normal, Type I, and Type II materials in Altair Flux FEM software.

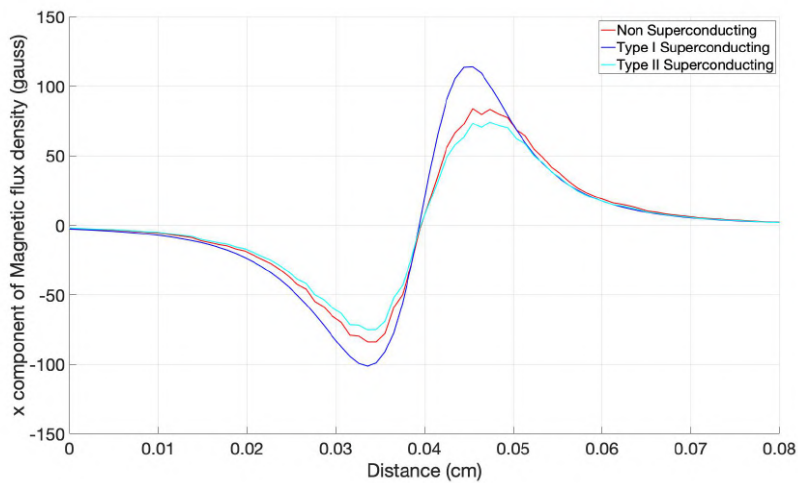

Fig. 21 x component of field vs distance

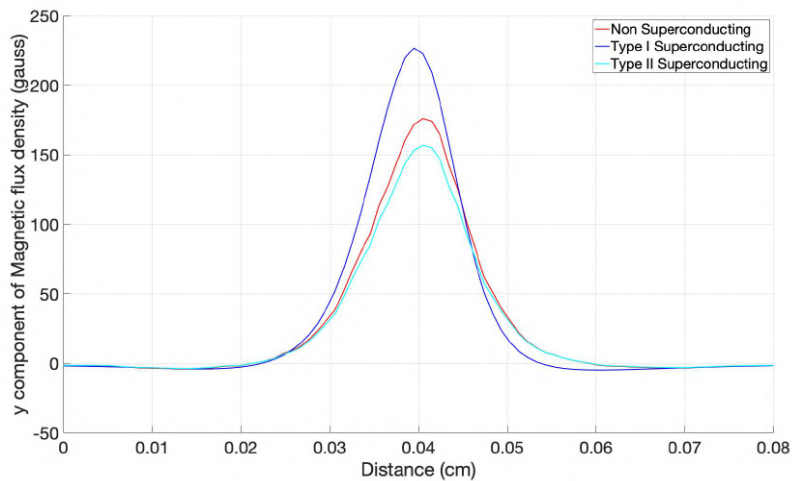

Fig. 22  y component of field vs distance

Fig. 23 and Fig. 24 show simulation results of x and y components of force vs distance curves respectively using normal, Type I, and Type II materials.

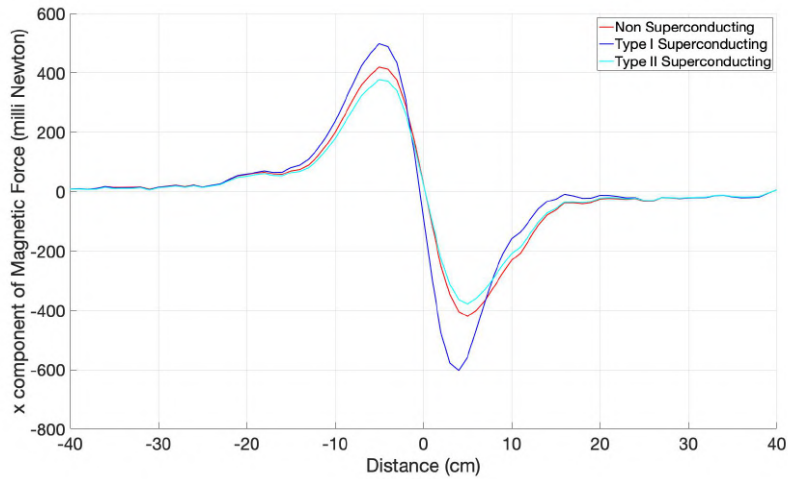

Fig. 23 x component of force vs distance

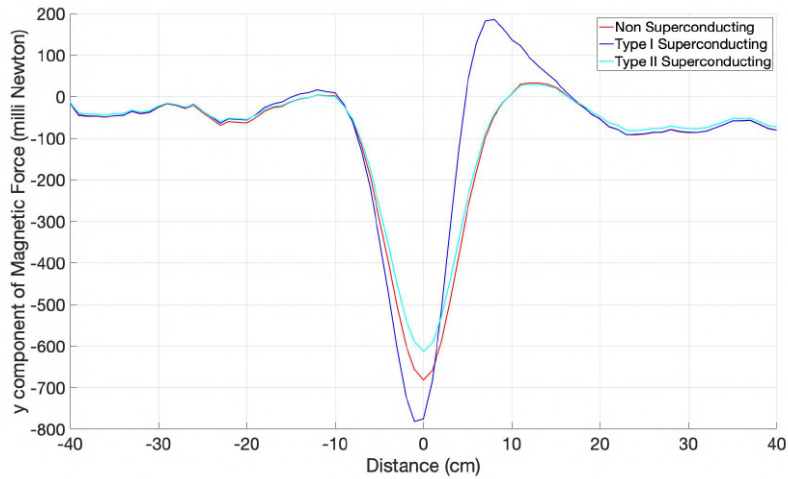

Fig. 24 y component of force vs distance

**6. Discussions:** From the experimental and theoretical results it is observed that the magnitude of peak field or force vary due to magnetisation of superconductor. All results are taken when the superconductor is maintained in the same state through out the distance. A net work done in moving the magnet from left to right can be realised by switching the superconducting state as shown in Fig. 25, hence switching between states E and F in Fig. 4 with quasi static approach.

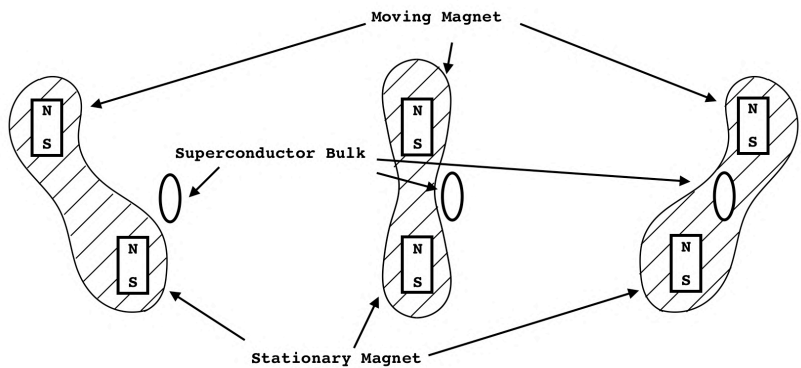

Fig. 25 Switching of Superconductivity for braking applications

**Patent Publications:** Superconducting Permanent Magnetic Motor, publication number 46/2017, date: 04/12/2015

**Acknowledgement:** The authors would like to thank Prof. Sethupathi and Prof. Sankaranarayanan, Dept. of Physics, IIT Madras, for providing the superconductor sample.